**Light-induced hysteresis of electronic polarization in antiferromagnet FePS$_3$**

*Kyung Ik Sim*[1,2,†], *Byung Cheol Park*[1,2,†], *Taesoo Kim*[1,2], *Byeong Wook Cho*[1,2], *Jae Hoon Kim*[3], *Eun-Mi Choi*[1,2,*], *and Young Hee Lee*[1,2,*]

[†] These authors equally contributed to this work.

K. I. Sim, B. C. Park, T. Kim, B. W. Cho, E.-M. Choi, Y. H. Lee
Center for Integrated Nanostructure Physics, Institute for Basic Science(IBS)
Sungkyunkwan University (SKKU)
Suwon 16419, Republic of Korea
E-mail: emchoi@skku.edu; leeyoung@skku.edu

J. H. Kim
Department of Physics, Yonsei University
Seoul 03722, Republic of Korea

**Abstract**
Research on manipulating materials using light has garnered significant interest, yet examples of controlling electronic polarization in magnetic materials remain scarce. Here, we demonstrate the hysteresis of electronic polarization in the antiferromagnetic semiconductor FePS$_3$ via light. Below the Néel temperature, we observe linear dichroism (i.e., optical anisotropy) without structural symmetry breaking. Light-induced net polarization aligns along the a-axis (zigzag direction) at 1.6 eV due to the dipolar polarization and along the b-axis (armchair direction) at 2.0 eV due to the combined effects of dipolar and octupolar polarizations, resulting from charge transfer from the armchair to the zigzag direction by light. Unexpected hysteresis of the electronic polarization occurs at 2.0 eV due to the octupolar polarization, in contrast to the absence of such hysteresis at 1.6 eV. We attribute this to a symmetry breaking of the light-induced phase of FePS$_3$ involving electronic polarization within the spin lattice. This study suggests a new mechanism for generating and controlling electronic polarization in magnetic materials using light, with implications for future device applications.

1. Introduction

The advent of van der Waals (vdW) materials has driven substantial advancements in science. With their weak interlayer interactions, these materials enable two-dimensional (2D) structures through mechanical exfoliation, leading to developments in valleytronics,[1–3] excitonics,[4–6] and superconductivity[7–9] via dimensional reduction, primarily in electronic effects. Beyond electronic properties, exotic magnetic effects have been observed in the 2D limit of vdW materials,[10] revealing 2D (anti)ferromagnetism and 2D multiferroicity,[11] challenging the conventional understanding of the Mermin-Wagner theorem, which posits that quantum fluctuations in 2D hinder the stabilization of an ordered magnetic ground state.[12]

Building on these advancements, researchers have sought additional control parameters to couple degrees of freedom in 2D vdW magnets,[13–15] complementing prior work focused on dimensional reduction. Light, in particular, has emerged as a powerful tool for manipulating coupling between degrees of freedom. For example, transient or metastable states with novel physical properties, unavailable in the ground state, have been demonstrated through ultrafast pump-probe techniques that allow dynamic control of such coupling.[16–19] Additionally, a light-induced memory effect in steady-state measurements enables modifications of the ground state itself.[20] Light couples directly with orbitals[21], especially in semiconductors, making light-orbital coupling a promising route for tuning the physical properties of semiconducting vdW magnets. Here, we focus on the potential of controlling electronic polarization via light-orbital coupling, a phenomenon that remains exceptionally rare.

We anticipate successful control of electronic polarization in semiconducting vdW magnets (multiferroicity), based on recent advances. For instance, light-induced electronic polarization has been reported in antiferromagnetic $Cr_2O_3$, induced by in-plane rotational symmetry breaking due to optical rectification.[22] Also, dynamical multiferroicity has been demonstrated in $SrTiO_3$ through terahertz electric field-driven lattice vibrations.[23] Efforts are also underway to establish single-layer multiferroic properties in vdW non-collinear antiferromagnet $NiI_2$[24] and non-vdW ferromagnet $CuCrSe_2$.[25] However, the realization of 2D multiferroicity remains debated due to the indirect nature of all-optical methods,[26] underscoring the need for direct observation of light-induced hysteresis of electronic polarization. Achieving optical control of electronic polarization and demonstrating its hysteresis in semiconducting vdW magnets could exceed the capabilities of conventional methods relying on electric and magnetic fields, thereby broadening the potential applications of light in material manipulation.

## 2. Results and Discussion

Here, we report the observation of light-induced hysteresis in electronic polarization for antiferromagnetic (AFM) semiconductor FePS$_3$ using confocal optical spectroscopy. Visible photons induce electronic polarization by promoting bound electrons across the band gap, distinct from conventional approaches using electric and magnetic fields only responsible for the bound electrons. The transmission spectrum (1.4–2.2 eV) and dielectric constants unveil the emerging optical anisotropy below the Néel temperature ($T_N$ ~117 K), providing direct evidence of light-induced electronic polarization. Broadband spectroscopy enables us to examine charge conservation across the spectral range using the optical sum rule, attributing the electronic polarization to the redistribution of charges across different energies. Our symmetry analysis indicates that the hysteresis in electronic polarization originates from mirror symmetry breaking in AFM materials, opening new avenues for realizing multiferroics by light.

**Figure 1**a shows the crystal structure of FePS$_3$ produced by VESTA.[27] The structural basis comprises hexagonal Fe atoms (red dashed line in the top figure) arranged in a honeycomb lattice, with each Fe atom coordinated by six S atoms (top figure) and the S atoms are positioned at the outmost of each layer (bottom figure). Depending on the crystal axis, Fe atoms form a different arrangement: along the **a**-axis (zigzag direction) and along the **b**-axis (armchair direction). The S atoms are bonded to two P atoms located above and below the Fe atomic plane (bottom figure), forming a vdW gap between each layer.[28]

Figures 1b,c present a schematic illustration of the experimental setup. Broadband (1.4–2.2 eV) light is incident to the sample (FePS$_3$ flakes on the quartz substrate), and the transmission (T) of light through the sample is measured by a confocal microscope. To isolate a single domain showing maximal transmission contrast, the optical beam was narrowed to a 5-μm diameter using an objective lens. Importantly, the polarization of incident light can be adjusted to be linearly polarized in parallel to the **a**-axis (zigzag) and **b**-axis (armchair). The transmission spectrum of FePS$_3$ (T$_{FePS3}$) is obtained by removing the substrate effect: the transmission of the entire sample T$_{sample\ (FePS3\ +\ sub.)}$ is normalized by that of the substrate T$_{sub}$ (see Supporting Information Figure S1 for the temperature dependence of T$_{FePS3}$). Subsequently, the transmission is transformed into absorbance (A), using the Beer-Lambert law, A = − ln(T).[29,30]

In Figure 1d, we present the absorbance spectrum of FePS$_3$ at 125 K (> $T_N$ ~117 K) using unpolarized light. Across the band gap (~1.4 eV) of semiconducting FePS$_3$, light triggers electronic charge excitation among $d$ orbitals of Fe$^{2+}$ ion, resulting in absorption.[31]

To pinpoint the absorption peak position, we analyze the 2nd derivative of the absorbance spectrum, which unveils a prominent absorption peak at ~1.7 eV (indicated by an orange triangle) corresponding to the *d-d* transition[31–35] (a schematic illustration in Figure 1e). Trigonal distortion in the octahedron allows the otherwise typically forbidden *d-d* transition.[31,36] The absorbance under unpolarized light is identical to that under linearly polarized light, indicating the isotropic nature of the optical response. Absorbance spectra along the **a**-direction and **b**-direction coincide, demonstrating the expected isotropy of the optical response above $T_N$, as expected from the honeycomb lattice.

However, below $T_N$, linear dichroism (LD) emerges as anisotropy without structural phase transition. In **Figure 2**a, we compare the directional absorbance spectra above and below $T_N$. At 10 K (below $T_N$), the absorbance in the **a**-direction does not coincide with that in the **b**-direction, in contrast to the 125 K case (above $T_N$). This anisotropy is accompanied by the Ising-type AFM spin ordering below $T_N$ (a schematic illustration in the inset).[36,37] Figure 2b presents temperature-dependent absorbance from 140 K to 10 K, showing the splitting of directional absorbance spectra across near $T_N$. To quantify the anisotropy below $T_N$, we calculate LD $\equiv (A_a - A_b) / (A_a + A_b)$, where $A_a$ and $A_b$ are the absorbances along **a** and **b** directions, respectively (Supporting Information Figure S2 for LD of four different samples). As shown in Figure 2c, the temperature-dependent LD values at 1.7 eV exhibit a first-order-like phase transition across $T_N$, from a paramagnetic to an AFM state, consistent with previous studies.[33–35] Note that the LD spectra in Figure 2c are normalized for the comparison of four samples (individual LD values for each sample are presented in Supporting Information Figure S2). Noticeably, no structural phase transition occurs across $T_N$, as confirmed by the temperature-dependent Raman data (Figure 2d).[37]

We turn our focus to the dielectric constant ($\varepsilon$) that provides information on the electronic polarization and joint density of states (JDOS). By fitting the experimental to the modeled transmission using dielectric function models,[38–41] we obtain the real (Re) and imaginary (Im) parts of the complex $\varepsilon$ (= Re[$\varepsilon$] + $i$*Im[$\varepsilon$]) (see Figure S3). We note that the real part of $\varepsilon$ is proportional to the electronic polarization whereas the imaginary part of $\varepsilon$ is proportional to JDOS, a product of the number of initial and final states.[29,30] This analysis reveals the process of electronic charge excitation, transitioning from the higher energy of the **b**-axis to the lower energy of the **a**-axis. In **Figure 3**a, at 125 K (above $T_N$), the Im[$\varepsilon$] values for both axes are nearly identical. This reflects isotropic JDOS in the optical transition above $T_N$ in FePS$_3$ (dashed line in middle panel). However, at 10 K (below $T_N$), the Im[$\varepsilon$] values for both axes are different: Im[$\varepsilon$] along the **b**-axis decreases significantly over the energy range of 1.6–2.1 eV, suggesting a reduced optical absorption near 2.0 eV. In contrast, the Im[ε] along **a**-axis shows an increase in the range of 1.5–2.0 eV, indicating an enhanced optical absorption near 1.6 eV. Supporting the sum rule for Im[ε], the reduction (−) in the **b**-axis is nearly identical to the increment (+) in the **a**-axis, revealing light-induced charge transfer from the **b**-axis near 2.0 eV to the **a**-axis near 1.6 eV in the presence of spin ordering. This variation in optical absorption aligns well with the change in JDOS across $T_N$ (middle panel)[33,42] and is further elucidated by the change in Im[ε], quantified as ΔIm[ε] = Im[ε](10 K) – Im[ε](125 K), depicted in the right panel of Figure 3a (see Supporting Information Figure S4 for temperature dependence). This light-induced charge transfer from the **b**-axis to the **a**-axis, facilitated by spin ordering, underscores the spin-charge coupling in $FePS_3$.

Our analysis extends to Re[ε], a parameter linearly proportional to electronic polarization **P**. As depicted in Figure 3b, significant increases in electronic polarization are observed below $T_N$ along both the **a**-axis and **b**-axis, manifesting at different energies—near 2.0 eV for the **b**-axis and 1.6 eV for the **a**-axis. This electronic polarization enhancement is quantitatively represented by ΔRe[ε] = Re[ε](10 K) – Re[ε](125 K), illustrated in the right panel and detailed across all temperatures in Supporting Information Figure S4. Such variations imply that net electronic polarization aligns with the **b**-axis (**P** ∥ **b**) near 2.0 eV and with the **a**-axis (**P** ∥ **a**) near 1.6 eV, validating light-induced electronic charge excitation below $T_N$.

We explain the mechanism of light-induced charge excitation below $T_N$. The optically-induced intersite *d-d* transition (or electron hopping) occurs between $Fe^{2+}$ ions on the honeycomb lattice structure. This structure imparts directionality to this transition in real space: not only the zigzag direction (along the **a**-axis) but also armchair direction (along the **b**-axis) directions. Below $T_N$, the Ising-type AFM ordering further introduces spin-dependence into the *d-d* transition. The resulting light-induced electronic charge excitation takes place between $Fe^{2+}$ atoms with the same (opposite) spin configuration along the **a**(**b**)-direction (as described in Figure 3c,d). The light-induced charge excitation in the **a**-direction (zigzag direction, at ~1.6 eV) is energetically more stable than in the **b**-direction (armchair direction, at ~2.0 eV).

We present the *d*-orbital configuration of $FePS_3$, which is associated with our observation of optical *d-d* transitions. In the equilibrium, the orbital configuration is mainly determined by the Coulomb interaction (U) and Hund's exchange energy (J), as described in Figure 3c,d.[42] The initial state consists of two $Fe^{2+}$ ions with a $d^6$ configuration (six electrons in the *d* orbital), while the final state involves one $Fe^{2+}$ ion with a $d^5$ configuration and another with a $d^7$ configuration.[42–44] Due to the two possible spin states (up and down), two distinct final states arise after light-induced electron excitation. The $Fe^{2+}$ ion that loses an electron ($d^5$ configuration) has different energy levels depending on the spin orientation of the remaining electrons: all five spins aligned in the **a**-direction or one of the five spins opposite in the **b**-direction. While the $Fe^{2+}$ ion that gains an electron ($d^7$ configuration with five identical spins and two opposite spins) has the same energy level in both directions, the two distinct $d^5$ states lead to a splitting of the electronic polarizations: **P$_a$** in the **a**-direction (U–J ~1.6 eV) and **P$_b$** in the **b**-direction (U+J ~2.0 eV).

We next investigate ε(θ), the angle (θ)-dependence of the dielectric constant ε, where θ denotes the direction of the electric field **E** relative to the **a**-axis. We then plot the real part of permittivity Re[ε](θ) subtracted by its minimum value min(Re[ε](θ)), ranging from 0° to 90°, and its mirror range from 90° to 180° (see Figure S5 for angle dependence of optical spectra at 10 K and 125 K).

As a result, for 1.6 eV at 10 K (red dots and curve in **Figure 4**a), the ε(θ) data exhibits a decrease from 0° to 90° (bottom axis) for the Path I (black arrow), implying the strongest **P$_a$** in the **a**-direction and the weakest **P$_a$** in the **b**-direction. In the mirror range from 90° to 180° (top axis), the ε(θ) data shows a reversed trend, forming another path (Path II, green arrow). But Path II is identical to Path I for 1.6 eV at 10 K, implying the mirror symmetry of the entire system involving the light-induced **P$_a$**. This case resembles the situation observed at 1.6 eV and 125 K (grey dots and curve in Figure 4a), where the isotropic light-induced **P** makes the two paths indistinguishable due to the protected mirror symmetry. On the contrary, for 2.0 eV at 10 K, when **P** aligns with the **b**-axis, two paths now become split. This implies the breaking of mirror symmetry under the coexistence of spin ordering and **P$_b$**. At 10 K (blue dots and curve in Figure 4b), the ε(θ) data exhibits an increase from 0° to 90° and a decrease from 90° to 180° along the distinct path. A noticeable hysteresis gap, quantified by the maximum hysteresis gap (|δ$_{max}$|), is observed for 2.0 eV at 10 K, in contrast to the absence of a hysteresis gap for 1.6 eV at 10 K.

We emphasize that the hysteresis curve at 2.0 eV and energy-dependent |δ$_{max}$| are highly reproducible for three samples with different thicknesses, implying their inherent nature (Supporting Information Figure S6). To exclude potential extrinsic effects, we compare optical microscope images and transmission spectra of the sample immediately after exfoliation with those of the same sample after 9 months of air exposure. The optical microscope images reveal a clean surface without any bubble-like structures indicative of surface contamination and the nearly identical transmission spectra confirm the stability of the

physical properties (Supporting Information Figure S7). It is also reasonable to assume that surface contamination has a negligible impact on the bulk hysteresis, given the sample thickness (>100 nm) and the skin depth (~350 nm) of the incident light.

We next present the temperature dependence of the maximum hysteresis gap ($|\delta_{max}|$) in Figure 4c, revealing a strong correlation between hysteresis and spin ordering below $T_N$. The $|\delta_{max}|$ value is observed between $\theta = 45°$ and $135°$. Similar to the LD behavior (Figure 2c), the $|\delta_{max}|$ at 2.0 eV exhibits a first-order-like phase transition across $T_N$, which is fitted well by the BCS-type model, $|\delta_{max}|(T) = A*\tanh(2.3*((T_N/T)-1)^{1/2})$ where $T_N = 117$ K. In contrast, the $|\delta_{max}|$ at 1.6 eV remains nearly constant. Notably, the $|\delta_{max}|$ at 2.0 eV is also reproducible for three samples with different thicknesses, confirming the reliability of our data (Supporting Information Figure S8).

We propose that not only electronic dipoles but also higher-order multipoles, such as electronic octupoles, may contribute to the hysteresis gap. To elucidate the role of these multipoles, we perform a Fourier analysis of the angle-dependent real part of the dielectric constant ($Re[\varepsilon](\theta)$), revealing the presence of both dipolar ($\cos(2\theta)$) and octupolar ($\cos(6\theta)$) polarization components in Figure 4d (Supporting Information Figure S9). Indeed, the $Re[\varepsilon](\theta)$ exhibits two distinct behaviors at 1.6 eV and 2.0 eV. At 1.6 eV, the $Re[\varepsilon](\theta)$ can be well-fitted to a dipolar polarization model aligned along the **a**-axis. However, at 2.0 eV, the $Re[\varepsilon](\theta)$ deviates significantly from a pure dipolar model, necessitating the inclusion of an additional octupolar polarization component. We note that the dipolar polarizations at 1.6 and 2.0 eV are antiparallel, as experimentally confirmed by the phase analysis of the dipolar polarization (Supporting Information Figure S10). In contrast to the dipolar ($\ell = 1$) and octupolar ($\ell = 3$) polarizations, which break inversion symmetry, the quadrupolar polarization ($\ell = 2$) preserves inversion symmetry. Therefore, we have excluded the quadrupolar contribution from our Fourier analysis, as it is not directly related to the observed hysteresis.

We next attribute the electronic octupolar polarization to the origin of the hysteresis gap. In Figure 4e, we show the energy-dependent $|\delta_{max}|$. At lower energies from 1.4 to 1.7 eV, $|\delta_{max}|$ approaches zero, while at higher energies between 1.8 and 2.2 eV, it manifests a pronounced peak-like feature (Supporting Information Figure S11 for hysteresis curves at distinct energies). Surprisingly, the energy-dependent $|\delta_{max}|$ is consistent with the dispersion of the electronic octupolar polarization, as revealed by Fourier analysis of the angle-dependent $Re[\varepsilon](\theta)$ across various energies (Figure 4e). This contrasts with the dispersion of the electronic dipolar polarization, which is attributed to the origin of LD (inset of Figure 4e). Through Fourier analysis, we extract the contributions of electronic dipolar and octupolar polarizations to distinct physical properties: dipolar polarization for LD and octupolar polarization for $|\delta_{max}|$ (Supporting Information Figure S9). These findings underscore the crucial role of electronic multipolar polarizations in driving intriguing physical phenomena and warrant further investigation into their origins in related magnetic systems.

We also note that this profile of the energy-dependent $|\delta_{max}|$ is reproduced by the energy-dependent $\Delta Re[\varepsilon_b]$ along the **b**-axis (Supporting Information Figure S12). We find the direct proportionality between $|\delta_{max}|$ and $\Delta Re[\varepsilon_b]$ as follows: $|\delta_{max}| = \alpha \times \Delta Re[\varepsilon_b] + \beta$, where $\alpha =$ 0.13 and $\beta = 0.015$. This finding explicitly demonstrates the correlation between $|\delta_{max}|$ and $\mathbf{P_b}$. Comprehensively with the previous paragraph, this indicates that the combined effect of dipolar ($\mathbf{P_a}$) and octupolar polarizations results in the net polarization $\mathbf{P_b}$. There are two minor comments; i) $|\delta_{max}|$ is not reproduced by the energy-dependent $\Delta Re[\varepsilon_a]$ along the **a**-axis, implying the irrelevance of $\mathbf{P_a}$ to the mirror symmetry breaking we considered, and ii) $|\delta_{max}|$ values at 125 K are negligible, again affirming the isotropy of FePS$_3$.

Finally, we explain the hysteresis of light-induced electronic polarization with mirror symmetry breaking. Figures 4f-h show the basis honeycomb of Fe atoms in FePS$_3$ above $T_N$ (Figure 4f) and that below $T_N$ (Figures 4g,h). We note that the monoclinic crystal structure lacks mirror symmetry across both the **a**-axis ($M_a$) and **c**-axis ($M_c$). Consequently, considering the mirror symmetry across the **b**-axis ($M_b$) is sufficient to determine the inversion symmetry of the system, which is directly linked to the hysteretic behavior of the light-induced electronic polarization.

We will now discuss the **b**-axis in more detail. Above $T_N$, over the energies, the mirror symmetry is preserved across a mirror plane ($M_b$) perpendicular to the **b**-axis owing to the isotropic $\mathbf{P}$. Below $T_N$, the situation becomes more complicated due to the other origins, such as spin and anisotropic $\mathbf{P}$, for mirror symmetry breaking: both the spin and $\mathbf{P}$ flip by a mirror operation (Supporting Information Figure S13). Despite the AFM state of inherent spin ordering, at 1.6 eV (Figure 4e), the mirror symmetry is still preserved across a mirror plane ($M_b$), high-symmetric light-excited state, resulting from the equivalence between the original and mirror image (Figure 4e). Because of its mirror symmetry, the high-symmetric light-excited state, accompanied by $\mathbf{P_a}$ on the AFM spin lattice, has the mirror symmetry of dielectric constant with respect to $M_b$. Thus, the magnitude of $\mathbf{P_a}$ over $0° \leq \theta \leq 90°$ is naturally identical to that over its mirror range ($90° \leq \theta \leq 180°$), resulting in their overlapping paths (Path I = Path II).

Interestingly, at 2.0 eV (Figure 4h) below $T_N$—where $\mathbf{P_b}$ additionally breaks the mirror symmetry with respect to $M_b$, low-symmetric light-excited state. The $\mathbf{P_b}$ is a superposition of the electronic dipolar polarization $\mathbf{P_a}$ and the octupolar polarization. We again note that the $\mathbf{P_a}$ at 2.0 eV is still aligned with the **a**-axis, but the sign is inverted, as determined by phase analysis (Supporting Information Figure S10). This symmetry breaking in the low-symmetric light-excited state leads to a dielectric constant with broken mirror symmetry to $M_\mathbf{b}$. Therefore, the magnitude of $\mathbf{P_b}$ over $0° \leq \theta \leq 90°$ does not need to be identical to that over its mirror range ($90° \leq \theta \leq 180°$). This symmetry breaking allows different paths (Path I ≠ Path II), resulting in the hysteresis of electronic polarization as a highlight of this work. We provide additional information on energy-dependent hysteresis (Supporting Information Figure S11). The demonstration of the light-induced electronic polarization and its hysteresis along two distinct paths could be a promising route for creating multiferroics.

3. Conclusion

In this work, we demonstrate light-induced hysteresis in the electronic polarization of the antiferromagnetic vdW semiconductor $FePS_3$. Based on our rigorous analysis, we find that the electronic polarization in $FePS_3$ below the Néel temperature arises from a combination of dipolar and octupolar contributions. While the dipolar polarization contributes to linear dichroism, the octupolar polarization is the primary driver of the observed hysteresis. The interplay between spin ordering and electronic octupolar polarization, which is closely linked to charge transfer (spectral weight transfer) from the **b**-axis to the **a**-axis (Figure 3), is a unique feature of the low-temperature phase. This is in contrast to the situation above the Néel temperature, where only background isotropic polarization is present.

Our findings suggest that the light-induced electronic octupolar polarization, coupled with spin ordering, breaks inversion symmetry, leading to light-induced multiferroicity in semiconducting $FePS_3$. The emergence of hysteresis in the electronic polarization at 2.0 eV, but not at 1.6 eV, highlights the crucial role of symmetry breaking induced by the light-induced electronic octupolar polarization, which is present only near 2.0 eV. This symmetry breaking enables the existence of two distinct response paths required for the formation of hysteresis. Our findings show that light has the potential to be a very effective tool for creating and modifying the physical characteristics of semiconducting vdW magnets.

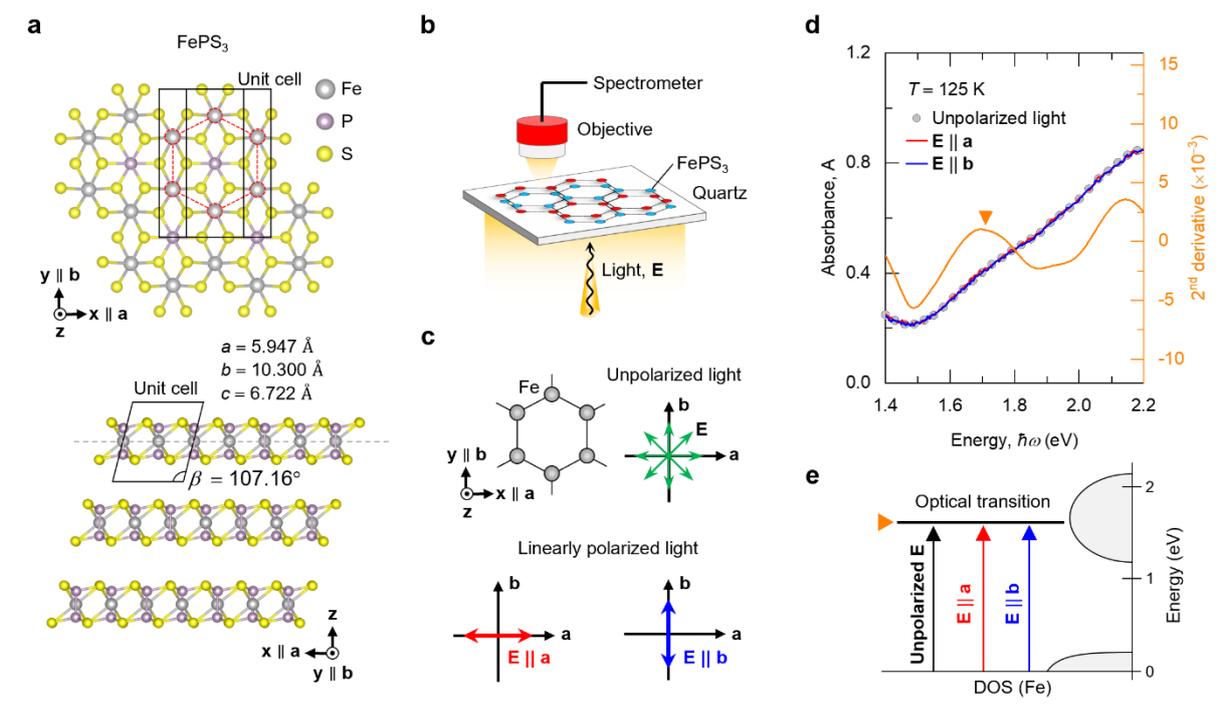

**Figure 1. Structural and optical properties of vdW semiconductor FePS$_3$. a**, Crystal structure of FePS$_3$ in the **a**-**b** plane (top) and **a**-**z** plane (bottom).[27,28] The red dashed line represents a honeycomb of Fe atoms. **b**, Experimental setup for confocal optical microscopy. A curly arrow with yellow shading indicates the incident light. **c**, Polarization (**E**) configuration of the light source. Unpolarized light (green) and linearly polarized light (red for **E** ∥ **a** and blue for **E** ∥ **b**) shining on the honeycomb of Fe$^{2+}$ ions (grey circles). **d**, Absorbance spectra of FePS$_3$ at 125 K (above $T_N$) with illumination by the unpolarized light (grey dots), **E** ∥ **a** (red line), and **E** ∥ **b** (blue line). Their second derivative curves (orange line) are also plotted. An orange triangle indicates the optical *d-d* transition peak. **e**, A simplified schematic illustration of the density of states (DOS) for Fe atoms, responsible for the optical *d-d* transitions.

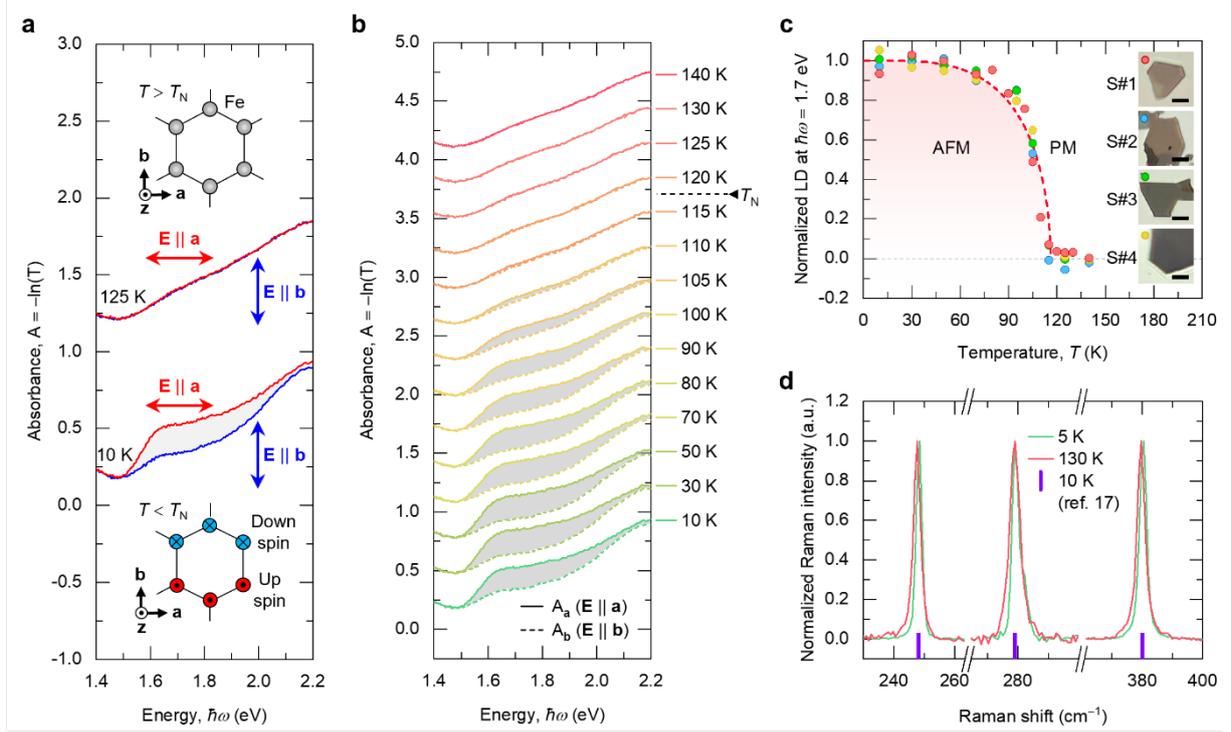

**Figure 2. Linear dichroism in FePS$_3$ below $T_N$. a**, Absorbance spectra for polarized light **E** || **a** (red) and **E** || **b** (blue) directions. The absorbance (A) is calculated by Beer-Lambert law A = −ln (T) from transmission T. The paramagnetic state (top inset figure) at 125 K (above $T_N$) is compared with the antiferromagnetic state (bottom inset figure) at 10 K (below $T_N$). **b**, Temperature-dependent absorbance spectra A. Solid lines represent absorbance $A_a$ for **E** || **a**, whereas dashed lines show absorbance spectra $A_b$ for **E** || **b**. Grey regions present the splitting between $A_a$ and $A_b$. **c**, Temperature-dependent linear dichroism (LD) at 1.7 eV for four different samples. Here, LD is defined by $(A_a - A_b)/(A_a + A_b)$, and its magnitude is normalized for direct comparison. The dashed curve is a guide to the eyes. The pictures on the right show optical microscope images of four samples. The scale bar is 10 μm. **d**, Temperature-dependent Raman spectra involving three phonon modes. No clear shift of the Raman peak is observed between 5 K (below $T_N$, green) and 130 K (above $T_N$, red). The violet ticks are the phonon frequencies at 10 K in the reference paper.[37]

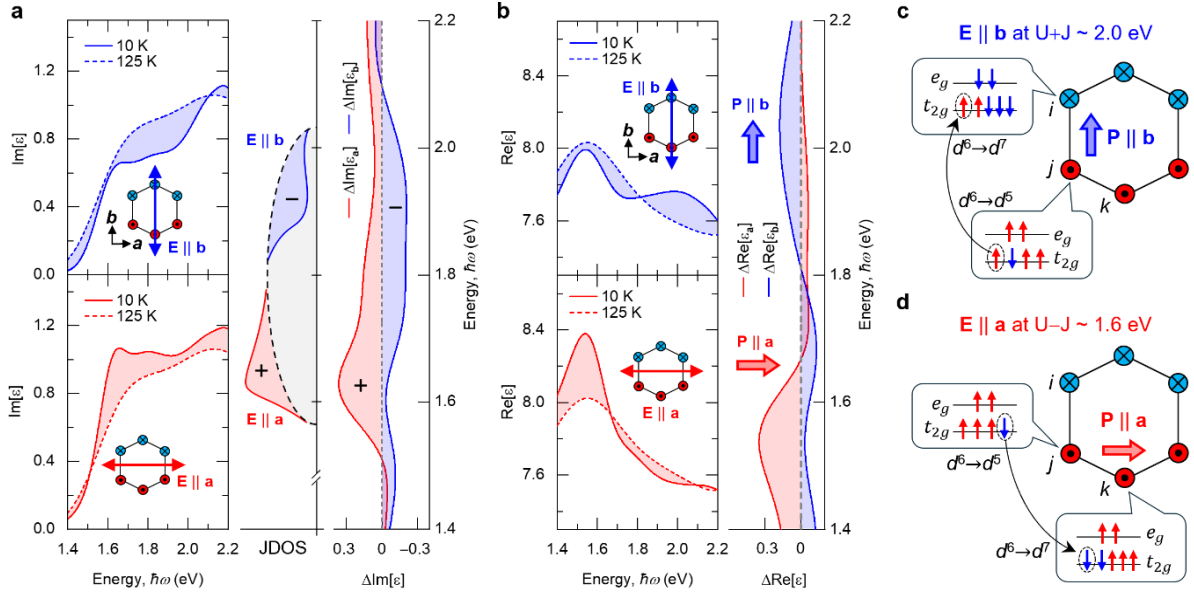

**Figure 3. Light-induced electronic polarization in AFM FePS$_3$ below $T_N$. a**, The imaginary part of the dielectric constant, Im[$\varepsilon$], for **E** ∥ **b** (left top, blue) and **E** ∥ **a** (left bottom, red) at 10 K (solid line) and 125 K (dashed line). Inset figures describe the polarization (double-sided arrow) of light. The estimation of the joint density of states (JDOS, middle panel) from the Im[$\varepsilon$]. The dashed line represents the JDOS at above $T_N$ whereas the solid line indicates the JDOS change below $T_N$. + represents the increase of JDOS at ~1.6 eV (red line) and – represents the decrease of JDOS at ~2.0 eV (blue line). The right panel depicts ΔIm[$\varepsilon$] = Im[$\varepsilon$](10 K) – Im[$\varepsilon$](125 K), where + and – signs represent an increase and decrease in absorption, respectively. Dots are experimental data whereas lines are guides to the eyes. **b**, The real part of the dielectric constant, Re[$\varepsilon$], for **E** ∥ **b** (left top, blue) and **E** ∥ **a** (left bottom, red) at 10 K (solid line) and 125 K (dashed line). In the right panel, ΔRe[$\varepsilon$] shows a peak at ~1.6 eV (red) and another peak at ~2.0 eV (blue). The light-induced electronic polarizations are described by the schematic illustrations (bold arrows, blue for **P$_b$** and red for **P$_a$**). Dots are experimental data whereas lines are guides to the eyes. **c**, The light-induced charge transfer between Fe$^{2+}$ ions along the **b**-direction occurs at U+J ~ 2.0 eV. This intersite $d$-$d$ transition involves Fe$^{2+}$ ions with opposite spins, resulting in a net electronic polarization (bold blue arrow) along the **b**-axis. The initial ground state consists of two Fe$^{2+}$ ions with a $d^6$ configuration (six electrons in the $d$ orbitals), while the excited state involves one Fe$^{2+}$ ion with a $d^5$ configuration and another with a $d^7$ configuration. The curved black arrow indicates the light-induced transfer of a spin-up electron (dashed circles). The orbital configuration, including the $e_g$ and $t_{2g}$ orbitals, is determined by Coulomb repulsion (U) and Hund's exchange energy (J). **d**, The light-induced charge transfer between Fe$^{2+}$ ions along the a-direction occurs at U–J ~ 1.6 eV, resulting in a net electronic polarization (bold red arrow) along the a-axis. In this case, the light-induced transfer involves spin-down electrons (dashed circles).

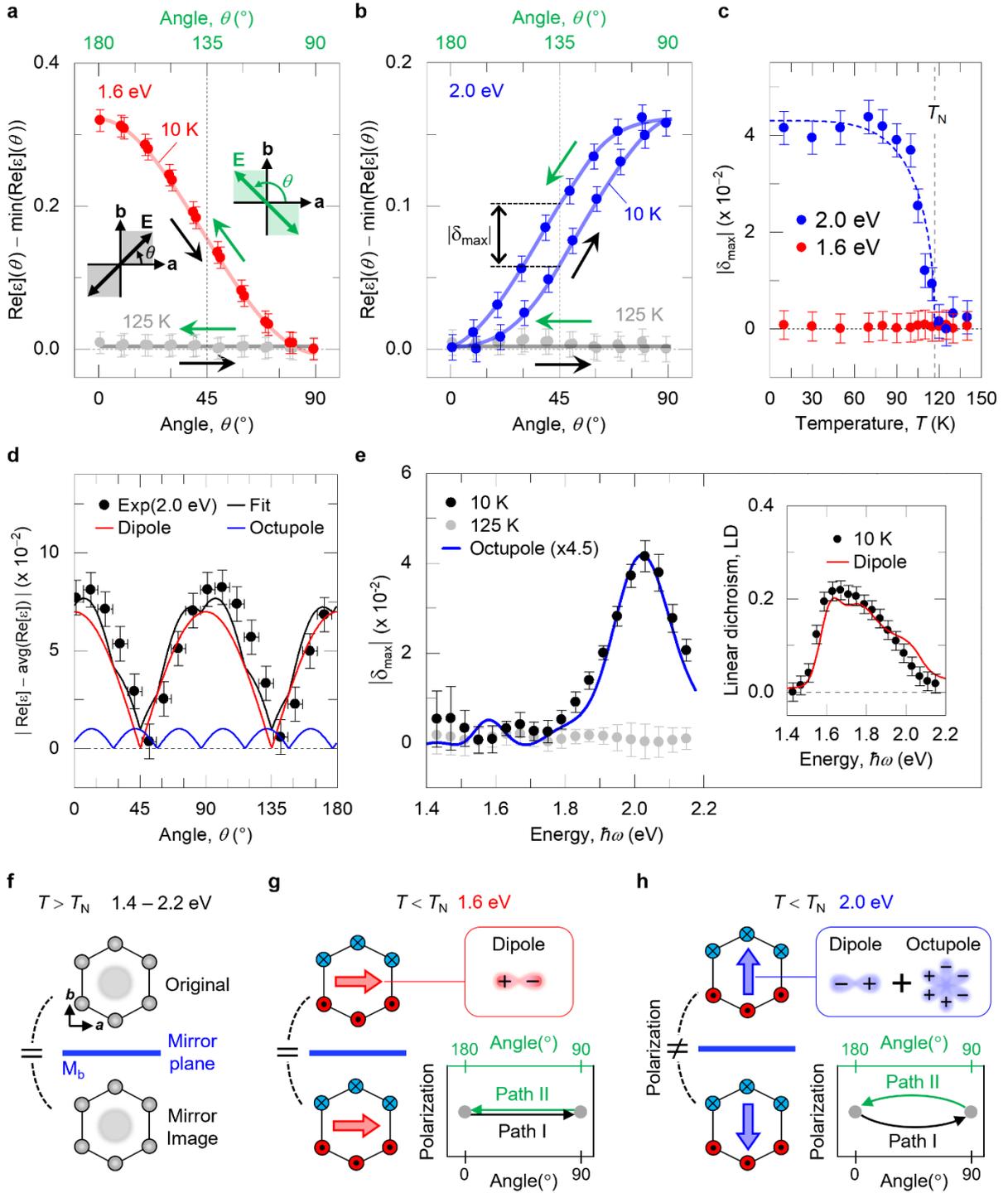

**Figure 4. Hysteresis of the light-induced electronic polarization in AFM FePS$_3$ below $T_N$.**
**a**, $\theta$-dependent $Re[\varepsilon](\theta) - \min(Re[\varepsilon](\theta))$ at ~1.6 eV at 10 K (below $T_N$, red dots and line). As illustrated in the inset figure, $\theta$ is the angle of **E** relative to the +**a**-direction over 0° to 180°. $\theta$-dependent $Re[\varepsilon](\theta) - \min(Re[\varepsilon](\theta))$ at ~1.6 eV at 125 K (above $T_N$, grey dots and line). **b**, $\theta$-dependent $Re[\varepsilon](\theta) - \min(Re[\varepsilon](\theta))$ at ~2.0 eV at 10 K (below $T_N$, blue dots and line). $Re[\varepsilon](\theta) - \min(Re[\varepsilon](\theta))$ at ~2.0 eV at 125 K (above $T_N$, grey dots and line). $|\delta_{max}|$ is the maximum hysteresis gap between 45° and 135° (double-sided arrow). The error bar in (a) and (b) is the experimental variation for eighteen times repeated measurements. **c**, The temperature-dependent $|\delta_{max}|$. At 2.0 eV (blue dots), this shows the first-order-like transition around $T_N$ while at 1.6 eV (red dots) no such transition occurs. Blue dashed

line shows the BCS-type model fitting results with $|\delta_{max}|(T) = A*\tanh(2.3*((T_N/T)–1)^{1/2})$ where A is 0.043 and $T_N$ = 117 K. **d**, Electronic multipole analysis of the angle-dependent electronic polarization. The Re[ε] – avg (Re[ε]) (black dots) is fitted with a summation (black curve) of electronic dipolar polarization (red curve) and octupolar polarization (blue curve). **e**, The energy-dependent $|\delta_{max}|$ at 10 K (black), in comparison with that at 125 K (grey) in the main panel. The energy dependent $|\delta_{max}|$ at 10 K (black) is consistent with the energy dependent magnitude of the octupolar polarization (blue line). Inset shows the linear dichroism spectra (black dots) follows the energy-dependent magnitude of electronic dipolar polarization (red line). **f-h**, Symmetry analysis for the origin of the hysteresis. Blue lines are mirror planes $M_b$ perpendicular to the **b**-axis. Mirror symmetry by those mirror plane is represented by "=" for symmetry preserved and "≠" for symmetry breaking. Above $T_N$ over 1.4–2.2 eV (**f**), the symmetry is preserved for $M_b$. Below $T_N$ at 1.6 eV (**g**), the figure shows the preserved mirror symmetry for $M_b$ in the high-symmetric light-induced state (the spin ordering below $T_N$ together with light-induced electronic dipolar polarization $P_a$). The magnitude of $P_a$ over 0° ≤ θ ≤ 90° is identical to that over its mirror range (90° ≤ θ ≤ 180°), leading to their overlapped paths (Path I = Path II). Below $T_N$ at 2.0 eV (**h**), where $P_b$ breaks the mirror symmetry with respect to $M_b$, low-symmetric light-excited state. The $P_b$ is comprised of electronic dipolar polarization and octupolar polarization. The magnitude of $P_b$ over 0° ≤ θ ≤ 90° does not need to be identical to that over its mirror range (90° ≤ θ ≤ 180°), resulting in the hysteresis of electronic polarization.

## 4. Experimental Section/Methods

*Sample preparation*

Quartz substrates, supplied by iNexus, INC., were initially cut to a size of 0.5 cm × 0.5 cm. To eliminate any unwanted organic residues and prepare the surface for subsequent material transfer, the substrates underwent sonication in acetone followed by isopropanol baths. $FePS_3$ flakes were then prepared from bulk crystals obtained from 2D Semiconductors using the mechanical exfoliation method, commonly referred to as the scotch tape technique. To ensure the removal of potentially oxidized surfaces, each flake was exfoliated repeatedly, at least ten times. The exfoliated $FePS_3$ flakes were carefully transferred onto the quartz substrates by pressing them with a cotton bud for one minute, a step crucial for achieving flakes with optimal thickness and adherence. The thickness of the transferred $FePS_3$ flakes was verified using atomic force microscopy (AFM, Park Systems, INC.). To mitigate any risk of oxidation or aging, all procedures related to the transfer and subsequent surface topography assessments were conducted inside a glovebox environment, maintaining $H_2O$ and $O_2$ levels below 1 ppm.

*Confocal microscope absorption spectroscopy*

Optical spectra measurements were conducted using a transmission geometry setup within an optical cryostat provided by Montana Instruments, closely following methodologies similar to those described in ref. 34. The experimental setup utilized a broadband white light source (OSL2IR, Thorlabs) coupled with a linear polarizer (WP25M-UB, Thorlabs) boasting an extinction ratio greater than 8000:1 across the spectral range of 480 nm to 4000 nm, ensuring the generation of linearly polarized light for our measurements. In our confocal spectroscopy configuration, a multi-mode fiber with a 50 μm core size channeled the light into a CCD spectrometer (CCS200, Thorlabs), which covers a spectral range of 200–1000 nm. This setup, combined with a 10× objective lens (M Plan Apo NIR, Mitutoyo) featuring a numerical aperture (NA) of 0.26, enabled the acquisition of optical spectra across 480–1800 nm with a focused sample area diameter of 5 μm. The transmission intensity of both the $FePS_3$ sample and the quartz substrate was measured, alongside a background signal obtained by blocking the light source. This allowed us to calculate the specific transmission for $FePS_3$ as $T = (T_{FePS3+Substrate} - T_{background})/(T_{Substrate} - T_{background})$, from which the absorbance spectra were derived using the relation $A = -\ln(T)$. To further decipher the real and imaginary parts of the optical constants, we applied a modeled transfer function—incorporating six Gaussian and one Cauchy model—to the experimental transmission data. This approach is detailed in the Methods section under "Modeled Transfer Function" and visualized in Supporting Information Figure S4. Notably, the modeled extinction coefficient ($k$) and the experimentally derived $k$ exhibited close agreement, particularly in spectral features such as peak positions and intensities, indicating the reliability of our fitting and modeling results.

*Modeled transfer function*

The transmission spectra were fitted to the modeled transmission by the transfer matrix method.[29,30,39] The resulting transmission $T_{FePS_3}$:

$$\frac{T_{FePS_3+Substrate}(\lambda)}{T_{Substrate}(\lambda)} = T_{FePS_3}(\lambda) = \left| \frac{\frac{2n_2}{n_2+n_1} \frac{n_3+n_1}{n_3+n_2} \exp\left(i\frac{2\pi}{\lambda}n_2 d\right)}{1 + \frac{n_3-n_2}{n_3+n_2} \frac{n_2-n_1}{n_2+n_1} \exp\left(i\frac{4\pi}{\lambda}n_2 d\right)} \right|^2$$

where $n_1$ (=1) is complex refractive index of vacuum, $n_2$ is complex refractive index of $FePS_3$, $n_3$ (1.54) is complex refractive index of quartz substrate, $\lambda$ is wavelength of incident light, $d$ (= 128 nm) is the thickness of $FePS_3$.

In this modeling, we introduce six Gaussian models and one Cauchy model as below (see Supporting Information Figure S3 for fitting results).

Gaussian model:[38]

$$\varepsilon_{Gaussian} = \varepsilon_{r,G} + i\varepsilon_{i,G}$$

$$\varepsilon_{i,G}(E) = A\left[\exp\left(-4\ln(2)\left(\frac{E-E_0}{\sigma}\right)^2\right) - \exp\left(-4\ln(2)\left(\frac{E+E_0}{\sigma}\right)^2\right)\right]$$

$$\varepsilon_{r,G}(E) = \frac{2A}{\sqrt{\pi}}\left[D\left(2\sqrt{\ln(2)}\frac{E+E_0}{\sigma}\right) - D\left(2\sqrt{\ln(2)}\frac{E-E_0}{\sigma}\right)\right]$$

where $A$ is resonance amplitude, $E$ is photon energy of the incident light, $E_0$ is resonance frequency, $\sigma$ is broadening parameter equal to full-width-half-maximum (FWHM), and $D$ stands for Dawson function $D(x) = \exp(-x^2)\int_0^x \exp(t^2)dt$.

Cauchy model:[39]

$$\varepsilon_{Cauchy} = \varepsilon_{r,C} + i\varepsilon_{i,C} = (n_C + ik_C)^2$$

$$n_C(\lambda) = \frac{A_C}{\lambda^2} + \frac{B_C}{\lambda^4}$$

$$k_C(\lambda) = \frac{C_C}{\lambda} + \frac{D_C}{\lambda^3} + \frac{E_C}{\lambda^5}$$

where $\lambda$ is the wavelength of the incident light, the constants from $A_C$ to $E_C$ are the dispersion parameter which describes the spectral dispersion at the low frequency affected by a certain resonance at higher frequency out of our measurement frequency range.

**Supporting Information**

Supporting Information is available from the Wiley Online Library or from the author.

Acknowledgements

This work was mainly supported by the Institute for Basic Science (IBS-R011-D1). B.C.P. acknowledges the financial support from NRF (2019R1A6A3A01096112). J.H.K. acknowledges the Samsung Science and Technology Foundation Grant (SSTF-BA2102-04) and Nano·Material Technology Development Program through the National Research Foundation of Korea(NRF) funded by the Ministry of Science and ICT (RS-2023-00281839).

Y.H.L., E.-M.C. & B.C.P. supervised this work. T.K. prepared the $FePS_3$ sample. K.I.S. conducted low-temperature confocal microscopic transmission measurements. B.W.C. performed Raman spectroscopic measurement. K.I.S., B.C.P. & J.H.K. analyzed and interpreted the experimental data. K.I.S., B.C.P., E.-M.C. & Y.H.L. wrote the manuscript with input from all authors.

Supporting Information for

**Light-induced hysteresis of electronic polarization in antiferromagnet FePS$_3$**

*Kyung Ik Sim*[1,2,†], *Byung Cheol Park*[1,2,†], *Taesoo Kim*[1,2], *Byeong Wook Cho*[1,2], *Jae Hoon Kim*[3], *Eun-Mi Choi*[1,2,*], *and Young Hee Lee*[1,2,*]

K. I. Sim, B. C. Park, T. Kim, B. W. Cho, E.-M. Choi, Y. H. Lee
Center for Integrated Nanostructure Physics, Institute for Basic Science(IBS)
Sungkyunkwan University (SKKU)
Suwon 16419, Republic of Korea
E-mail: emchoi@skku.edu; leeyoung@skku.edu

J. H. Kim
Department of Physics, Yonsei University
Seoul 03722, Republic of Korea

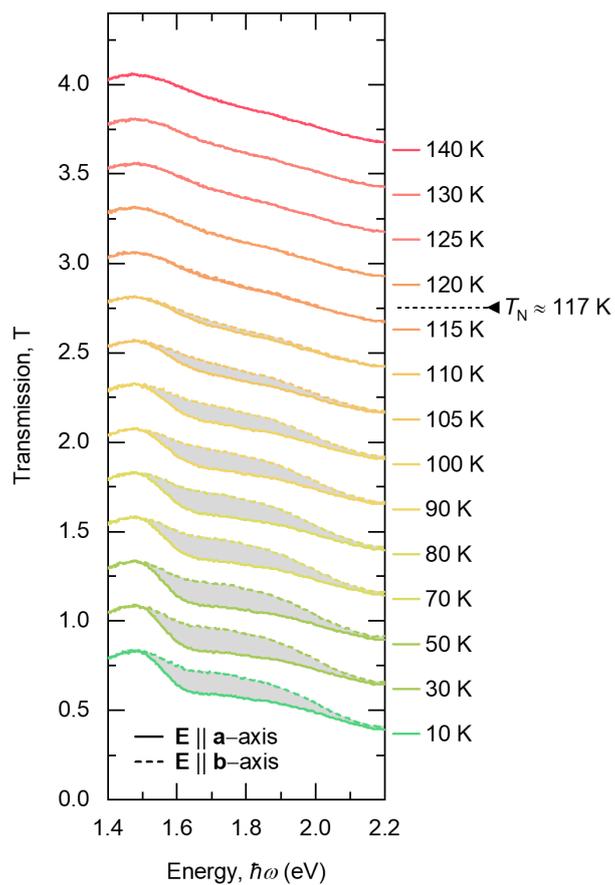

**Figure S1. Temperature-dependent transmission data of FePS$_3$.** Transmission spectra of FePS$_3$, normalized to the substrate transmission, are shown for linearly polarized light aligned with the **a**-axis (solid lines) and **b**-axis (dashed lines) across a range of temperatures.

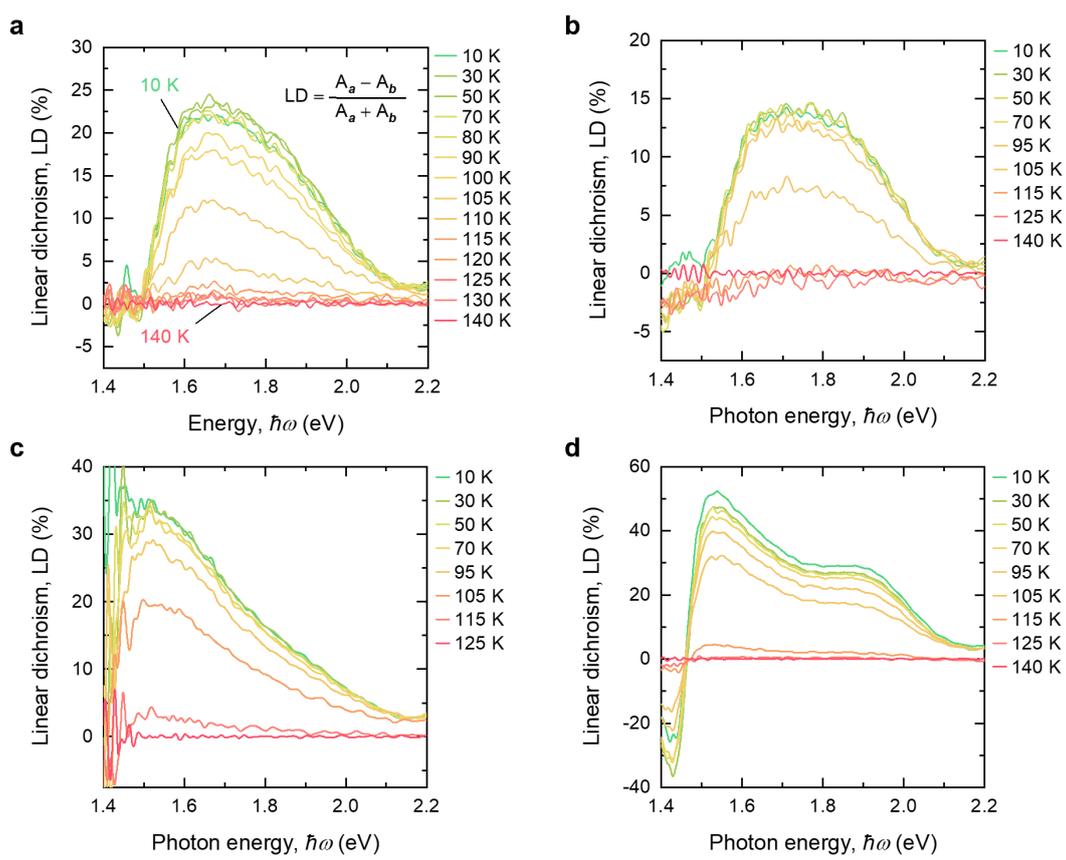

**Figure S2. Temperature and thickness dependencies of linear dichroism of FePS$_3$. a-d**, Temperature-dependent linear dichroism (LD) for four distinct FePS$_3$ samples of varying thicknesses: 128 nm (**a**, corresponding to red dots in Figure 2c of the main text), 123 nm, (**b**, blue dots) 156 nm (**c**, green dots), and 431 nm (**d**, yellow dots in Figure 2c). The differing spectral shapes of LD across these samples arise from internal reflection effects, the Fabry-Perot interference.

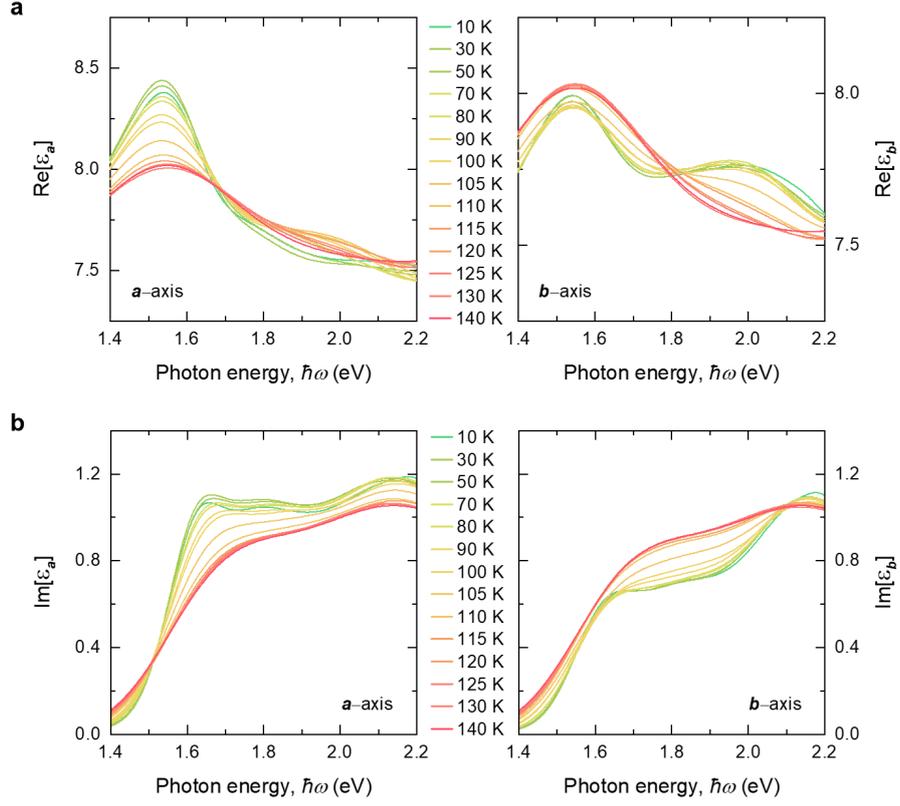

**Figure S3. Temperature-dependent dielectric constants of FePS$_3$ for *a*- and *b*-axes. a**, Real parts of the dielectric constant (Re[$\varepsilon$]) for the **a**-axis (left) and **b**-axis (right) across a temperature range from 10 K to 140 K. The data display a consistent temperature evolution with significant changes occurring below $T_N$. **b**, Imaginary parts of the dielectric constant (Im[$\varepsilon$]) for the **a**-axis (left) and **b**-axis (right) over the same temperature range.

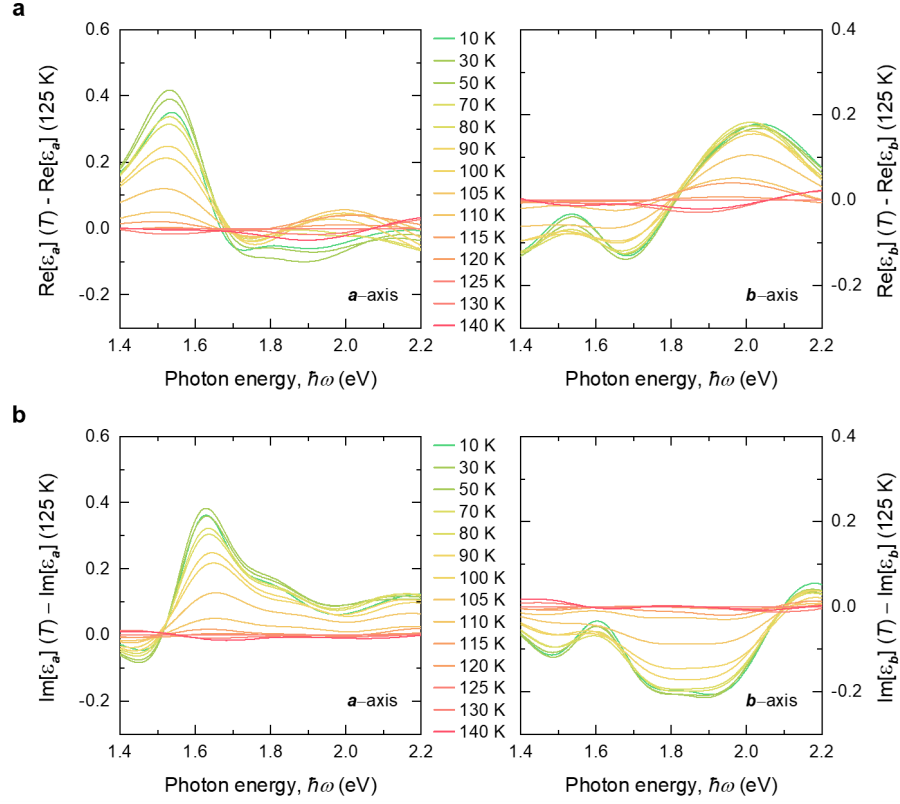

**Figure S4. Temperature-dependent change in dielectric constants of $FePS_3$. a**, Change in the real part of the dielectric constant, Re[$\varepsilon$], relative to its value at 125 K, plotted as Re[$\varepsilon$]($T$) − Re[$\varepsilon$](125 K) for the **a**-axis (left) and **b**-axis (right) across a temperature range from 10 K to 140 K. **b**, Corresponding changes in the imaginary part of the dielectric constant Im[$\varepsilon$], represented as Im[$\varepsilon$]($T$) − Im[$\varepsilon$](125$K$), for both axes over the same temperature range.

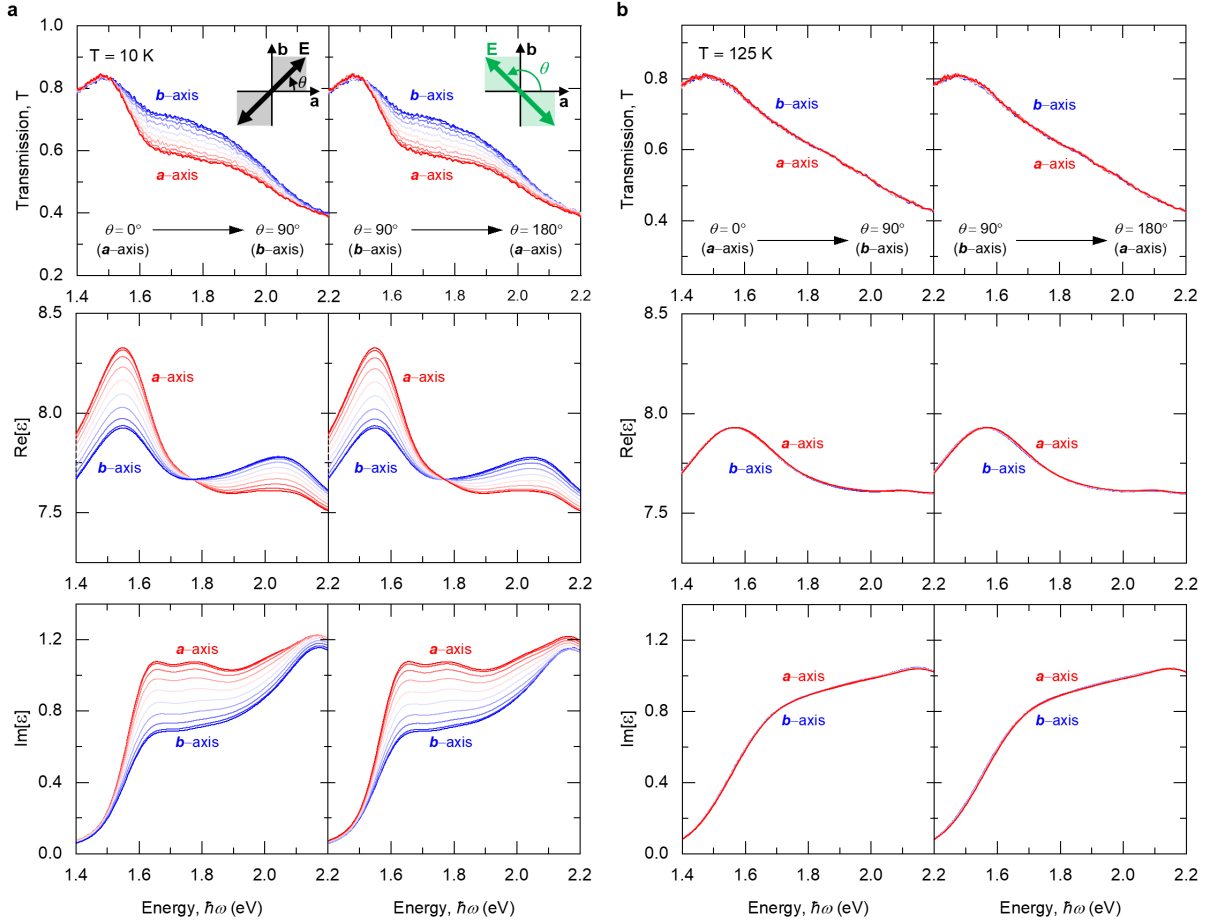

**Figure S5. Angle-dependent transmission and dielectric constants at 10 K and 125 K. a**, (Upper panel) Angle-dependent transmission data of FePS$_3$ at 10 K (below $T_N$) over the range of 0°–90° (left) and 90°–180° (right). (Middle panel) The angle-dependent real part of dielectric constants (Re[ε](θ)) at 10 K over the range of 0°–90° (left) and 90°–180° (right) degrees. (Bottom panel) The angle-dependent Imaginary part of dielectric constants Im[ε](θ) at 10 K over the range of 0°–90° (left) and 90°–180° (right) degrees. **b**, (Upper panel) The angle-dependent transmission data of FePS$_3$ at 125 K (above $T_N$) over the range of 0°–90° (left) and 90°–180° (right) degrees. (Middle panel) The angle-dependent real part of dielectric constants Re[ε] at 125 K over the range of 0° – 90° (left) and 90°–180° (right) degrees. (Bottom panel) The angle-dependent Imaginary part of dielectric constants (Im[ε](θ)) at 125 K over the range of 0°–90° (left) and 90°–180° (right) degrees.

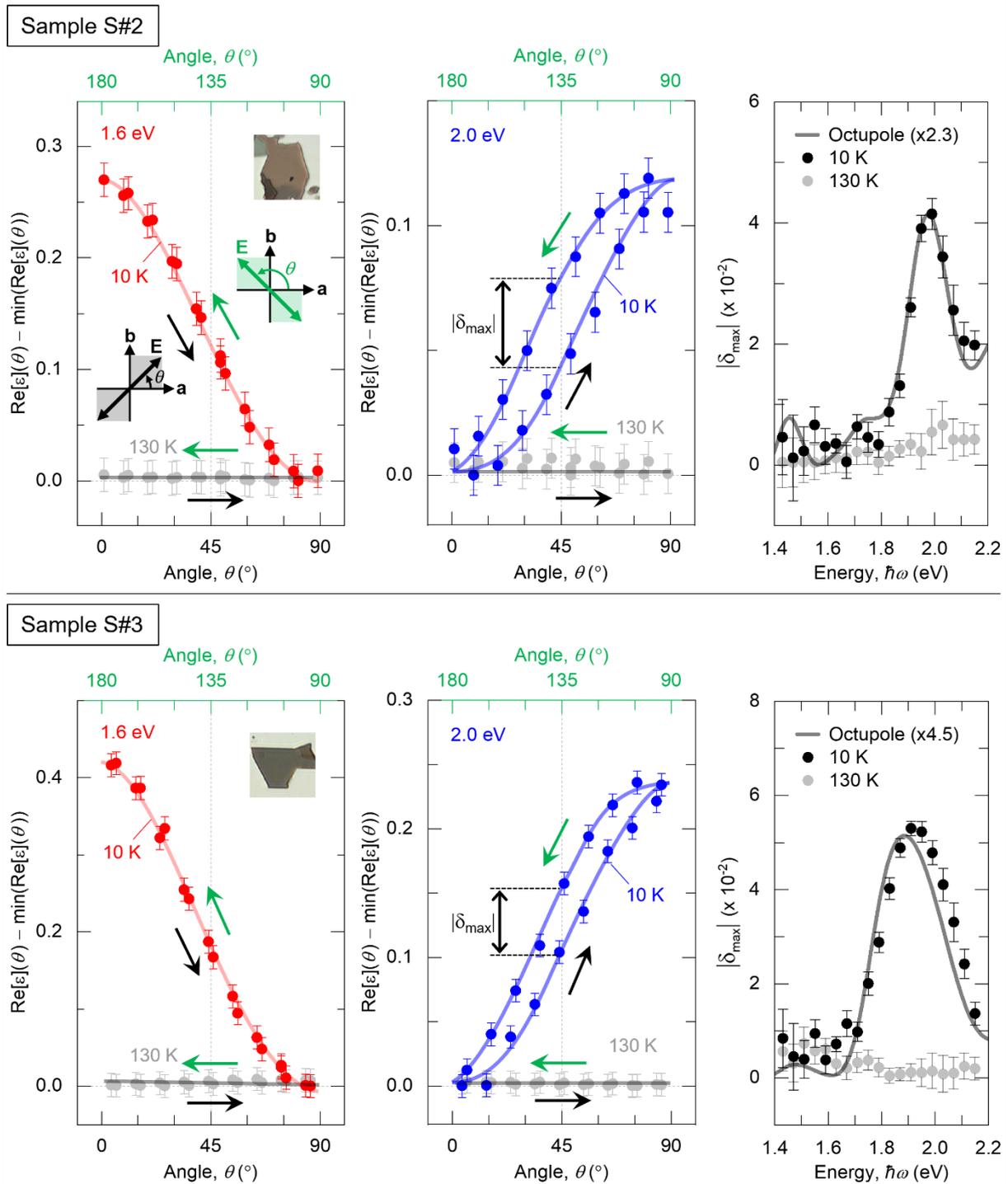

**Figure S6. Reproducibility of hysteresis for sample S#2 with a thickness of 123 nm (top panels) and for sample S#3 with a thickness of 156 nm (bottom panels).** For both samples, at 1.6 eV (left panel), angle-dependent Re[$\varepsilon$]($\theta$) - min(Re[$\varepsilon$]($\theta$)) shows the hysteresis without a gap |$\delta_{max}$| at 10 K (red), while no hysteresis appears at 130 K (grey). Dots are experimental data and lines are guides to the eyes. The Inset picture shows optical microscopic images of the sample. As illustrated in the inset figure, $\theta$ is the angle of E relative to the +a-direction over 0° to 180°. At 2.0 eV (middle panel), angle-dependent Re[$\varepsilon$]($\theta$) - min(Re[$\varepsilon$]($\theta$)) shows the hysteresis with a clear |$\delta_{max}$| at 10 K (blue), while no hysteresis appears at 130 K (grey). The energy-dependent |$\delta_{max}$| (right panel) shows its clear dispersion with a maximum peak near 2.0 eV at 10 K (black), while no dispersion appears at 130 K (grey). The experimental |$\delta_{max}$| (black dots) fits well with the dispersion of octupolar polarization (black line).

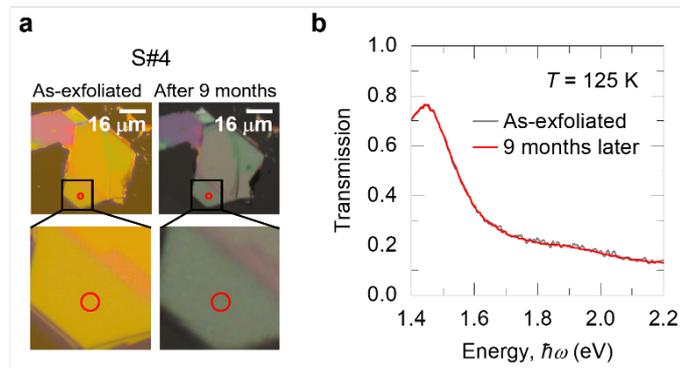

**Figure S7. Clean surface and absent aging effect. a, b,** Optical microscope images and transmission spectra of sample S#4, measured immediately after exfoliation and after 9 months of air exposure, respectively. The circles indicate the measurement positions, each with a beam spot diameter of ~5 μm.

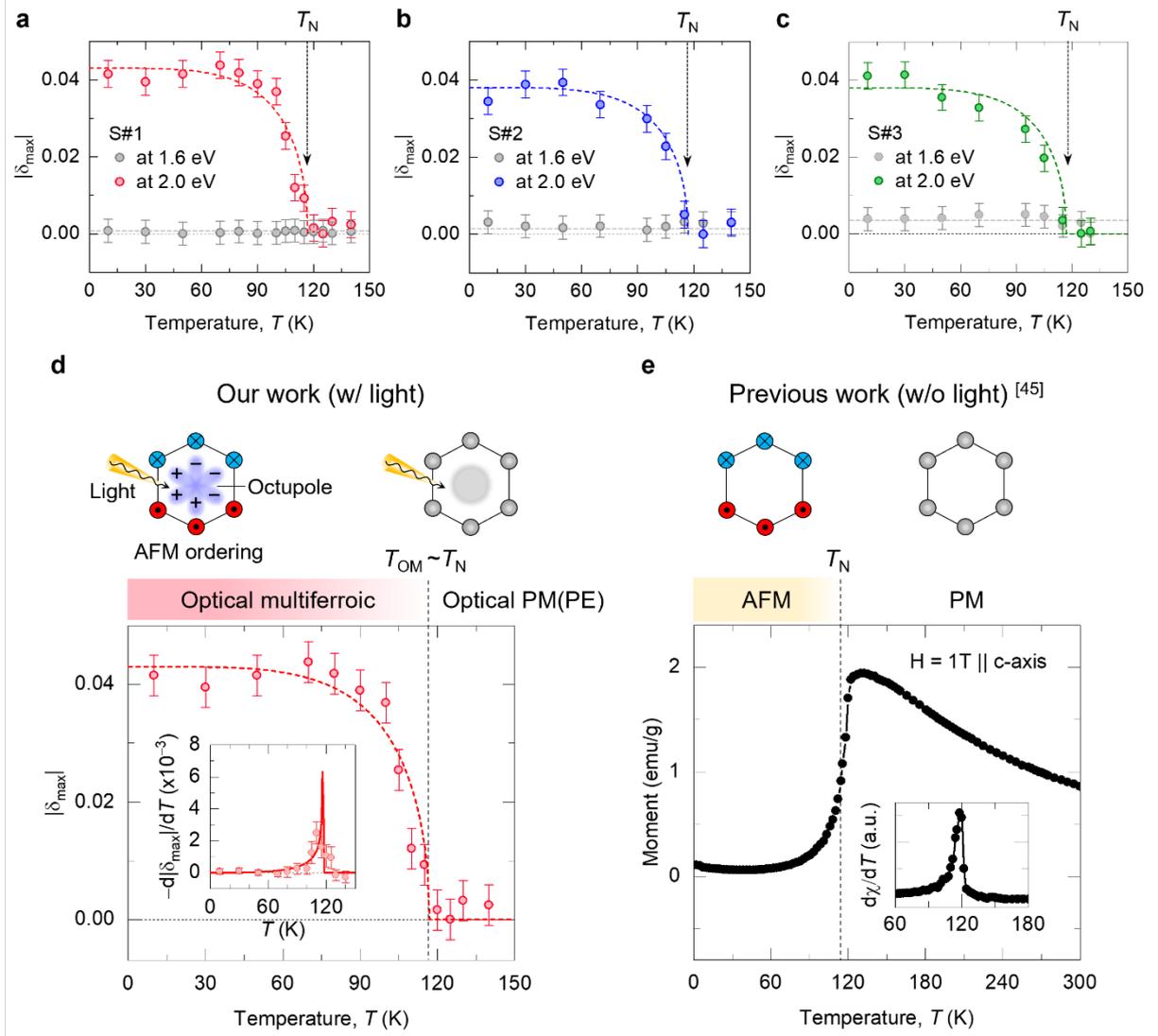

**Figure S8. Temperature-dependent hysteresis gap for the different samples. a-c,** temperature dependence of the maximum hysteresis gap ($|\delta_{max}|$) for samples S#1 (thickness $d$ = 128 nm), S#2 ($d$ = 123 nm), and S#3 ($d$ = 156 nm), respectively. For all samples, a first-order-like phase transition is observed in $|\delta_{max}|$ at 2.0 eV (colored dots), while no such transition occurs at 1.6 eV (grey dots). The dashed lines are BCS-type model fitting results with $|\delta_{max}|(T) = A*\tanh(2.3*((T_N/T)-1)^{1/2})$ where $T_N$ = 117 K. The amplitude A is 0.043 for S#1, 0.038 for S#2, and 0.038 for S#3. **d,** Phase diagram of light-excited FePS$_3$. The temperature dependence of $|\delta_{max}|$ for sample S#1 is shown in the main panel. The inset displays the derivative of $|\delta_{max}|$ with respect to temperature. A simple schematic of the phase is illustrated on the top. **e,** Phase diagram of unexcited FePS$_3$. The magnetic susceptibility data for unexcited FePS$_3$ is adapted from reference [45]. A simple schematic of the phase is illustrated on the top.

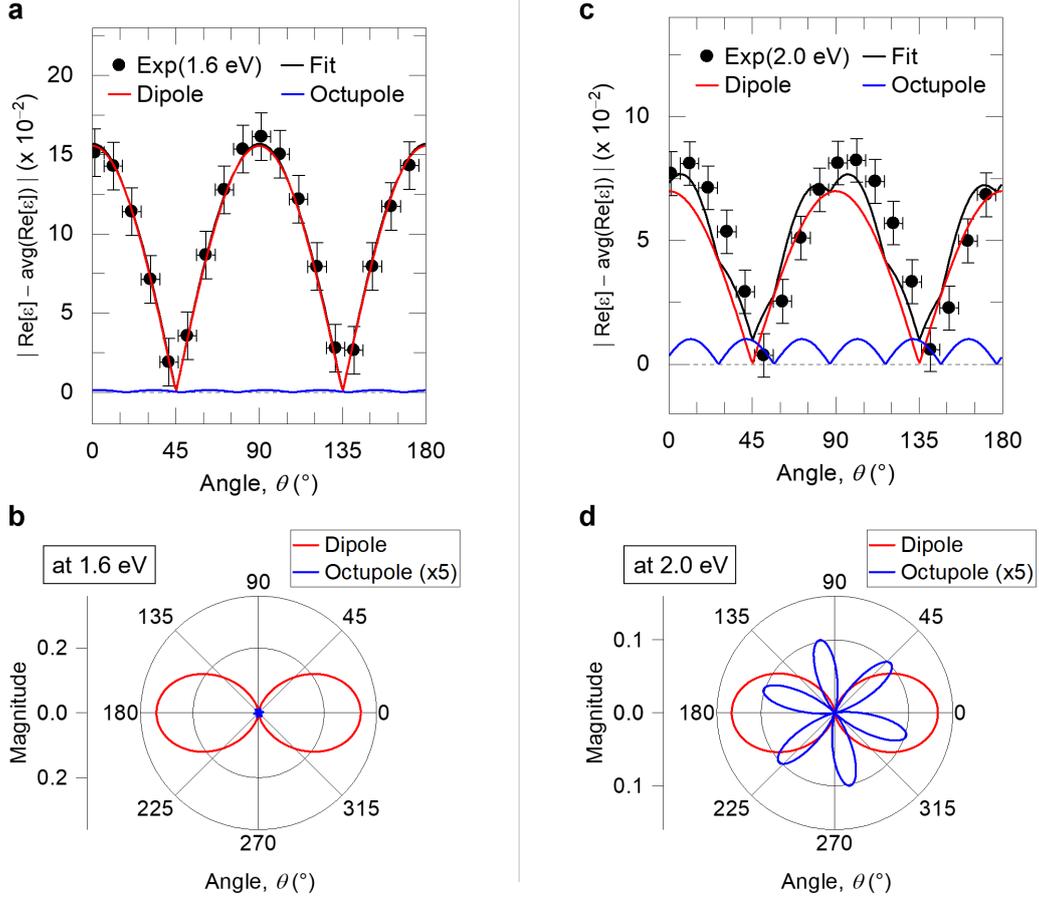

**Figure S9. Angle dependence of the real part of the dielectric constant due to dipolar and octupolar polarizations. a, b,** The angle-dependent dielectric constant at 1.6 eV, primarily influenced by electronic dipolar polarization along the **a**-axis. **c, d,** The angle-dependent dielectric constant at 2.0 eV, influenced by the electronic octupolar polarizations as well as the electronic dipolar polarization along the **a**-axis.

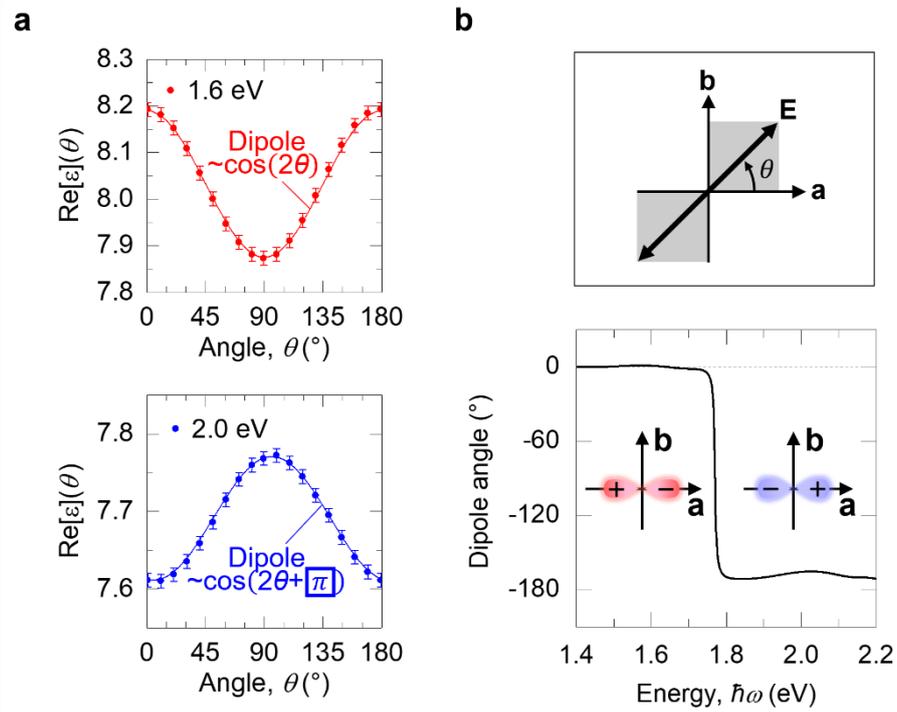

**Figure S10. Phase analysis of the electronic dipolar polarization ($P_a$) aligned to the a-axis. a,** Experimental data. Angle dependence of Re[ε] at 1.6 eV (red) and at 2.0 eV (blue). **b,** Calculation results, together with the schematic illustration. The phases of $P_a$ are inverted at 1.6 eV (red dumbbell) and 2.0 eV (blue dumbbell). The top panel describes the parameter (θ). **a** and **b** are the crystal axes and **E** is the polarization of light.

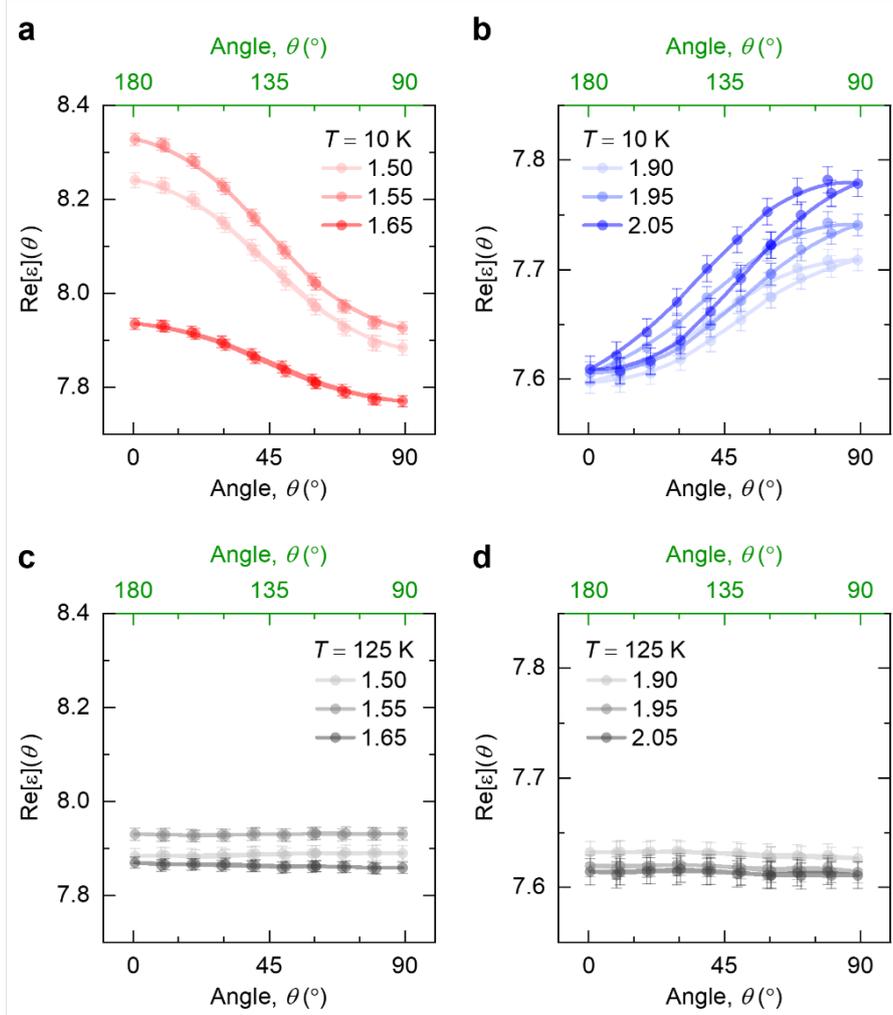

**Figure S11. The hysteresis of electronic polarization in FePS$_3$ for different photon energies at 10 K and 125 K. a**, The hysteresis of the real part of dielectric constant at 10 K, depending on the angle of electric field ***E*** along ***a*** direction. Red lines are hysteresis in the range of low energy part (1.50, 1.55, and 1.65 eV) dominated by electronic polarization **P$_a$** with no hysteresis gap at 10 K. **b**, Blue lines are the hysteresis of real part of dielectric constant in the range of high energy part (1.90, 1.95, and 2.05 eV) dominated by electronic polarization **P$_b$** with hysteresis gap at 10 K. **c-d,** Grey lines are the angle-dependent real part of dielectric constant at the energies of 1.50, 1.55, 1.65, 1.90, 1.95, and 2.05 eV at 125 K. All error bar is the experimental variation for eighteen times repeated measurement.

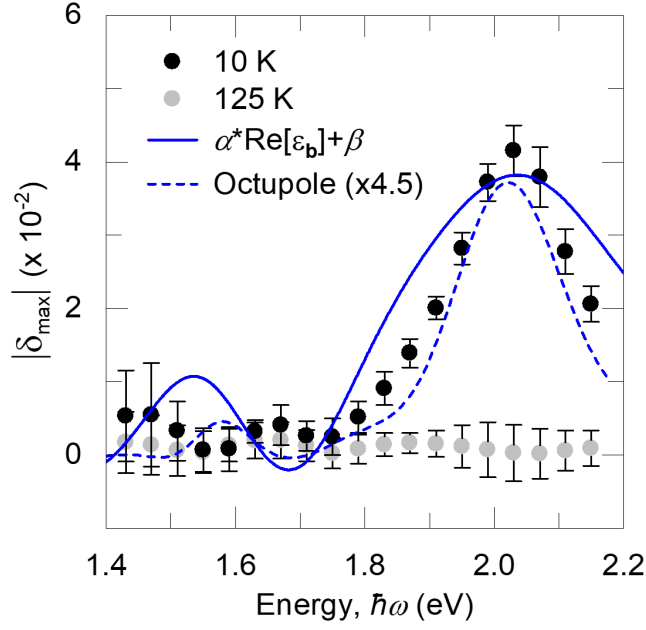

**Figure S12. Correlation among |δ$_{max}$(ω)|, Re[ε$_b$], and electronic octupole.** As discussed in the main text, the electronic octupolar polarization (blue dashed line) is responsible for the observed |δ$_{max}$| at 10 K (black dots). No significant |δ$_{max}$| is observed at 125 K (grey dots). The experimental |δ$_{max}$| at 10 K (black dots) aligns well with the theoretical curve (blue line) calculated using the relation |δ$_{max}$(ω)| = α × ΔRe[ε$_b$(ω)] + β, where α = 0.13 and β = 0.015. The temperature-dependent magnitude of the electronic octupolar polarization (blue dashed line) further confirms that the polarization along the **b**-axis (Re[ε$_b$], blue line) originates from the electronic octupolar polarization.

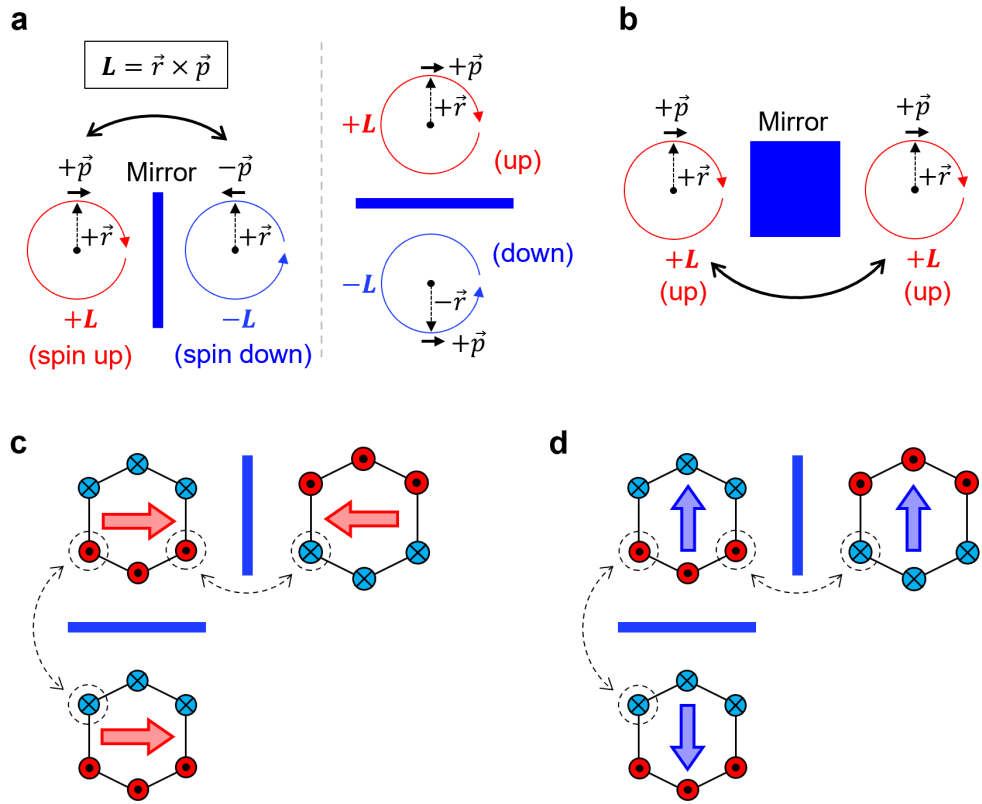

**Figure S13. Mirror symmetry operation and its implications for light-induced polarization in FePS$_3$.** In terms of symmetry, spin behaves similarly to angular momentum (L), defined as L = r × p, where r is the position vector and p is the momentum vector. Similarly, polarization (P) is defined as P = er, where e is the electronic charge. **a,b**, When a mirror plane is parallel to the direction of L, the mirror operation inverts the direction of L. However, when the mirror plane is perpendicular to L, the direction of L remains unchanged. **c,d**, When a mirror plane is perpendicular to the direction of P, the mirror operation inverts the direction of P. Conversely, when the mirror plane is parallel to P, the direction of P remains unchanged. **e,f**, In light-excited FePS$_3$, the combination of the optical *d-d* transition and Ising-type antiferromagnetic order (red and blue spins in the honeycomb lattice) below $T_N$ leads to a net polarization (bold arrows) within the honeycomb lattice.